\documentclass[showpacs,showkeys,11pt,
preprintnumbers,nofootinbib,
groupedaddress,superscriptaddress
]
{revtex4-1}

\usepackage[dvipdfmx]{graphicx}
\usepackage{cancel,subfigure}
\usepackage{color}
\usepackage{amsmath,amssymb}	
\usepackage{multirow}
\usepackage{enumerate}

\allowdisplaybreaks
\def\Dlr{\mathrel{\raise1.5ex\hbox{$\leftrightarrow$\kern-1em\lower1.5ex\hbox{$D$}}}}

\begin{document}
\title{Measurement of $A_{LR}$ using radiative return at ILC 250}
\preprint{}






\author{Takahiro Mizuno}
\affiliation{The Graduate University for Advanced Studies, SOKENDAI}
\author{Keisuke Fujii}
\affiliation{IPNS, KEK}
\author{Junping Tian}
\affiliation{ICEPP, The University of Tokyo}



\begin{abstract} 
\begin{center}
                     ABSTRACT
\end{center}
For the precision study at the ILC 250, measurement of $A_{LR}$ is important as it can constrain SMEFT parameters. 
The current best measured $A_{LR}$  value is $A_{LR} = 0.1514 \pm 0.0019 \, (stat) \pm 0.0011 \, (syst)$ which was measured at the SLC, and a more precise value is required for the global fit for the new physics search in TeV-scale.
At the ILC, we can use the $e^+ e^- \to \gamma Z$ process to evaluate the $A_{LR}$. 
We performed a full simulation study of the $e^+ e^- \to \gamma Z$ process at the center-of-mass energy of 250 GeV and evaluated how much we can improve the precision of this observable.
The statistical error on $A_{LR}$ at the ILC 250 turned out to be $1.8 \times 10^{-4}$.
Major source of the systematic error was error from the beam polarization.
As other sources of the systematic error, the uncorrelated parts of error on the product of luminosity and selection efficiency for each polarization combination contribute.
Including those systematic errors, total absolute error on $A_{LR}$ was estimated to be $0.00025$, $8.8$ times better precision than that from the SLC ($0.00219$).
\\
\\
\\
\\
\\
\\
\\
\\
\\
\\
\\
\\
\\
\\
\\
This is a preliminary study performed in the framework of the ILD concept group.
 \end{abstract} 
\maketitle

\section{Introduction}
\label{sec: Introduction}
At a polarized $e^+e^-$ collider, $A_e$ is given by the left-right asymmetry $A_{LR}$ in the total rate for $Z$ production,
\begin{equation}
A_e=A_{LR} \equiv \frac{\sigma_L-\sigma_R}{\sigma_L+\sigma_R},
\end{equation}
where $\sigma_L$ and $\sigma_R$ are the cross section for 100\% polarized $e^{-}_{L}e^{+}_{R}$ and $e^{-}_{R}e^{+}_{L}$ initial states. This $A_{LR}$ is important for the electroweak study, and it induces corrections to the $e^+e^-\to Zhh$, $e^+e^-\to Zh$, and $e^+e^-\to Z$ ($Z$-pole) processes.
Therefore, it can provide a very useful constraint for operators $c_{HL}$, $c_{HL}^{\prime}$, and $c_{HE}$ in the global SMEFT fit \cite{001}\cite{002}\cite{003}\cite{004}.

It turned out that the precision of the $A_{LR}$ measurement performed with the SLD detector at the SLAC Linear Collider (SLC), being at around 1.5\% \textit{i.e.}\,$A_{LR} = 0.1514 \pm 0.0019 \, (stat) \pm 0.0011 \, (syst)$\,\cite{lepslc}, is not precise enough for the global fit.
There were 2 dominant systematic errors in the measurement of $A_{LR}$ in the SLD: uncertainty of beam $E_{CM}$ and uncertainty of beam polarization.
At the ILC 250, we can use the radiative return process, $e^+e^-\to \gamma Z$, to measure the $A_{LR}$ and it has roughly 150 times more statistics than the SLC had.
There is a fast detector simulation study available for this reaction \cite{Alr}. Then we tried to perform full detector simulation study to get more realistic estimations including systematic errors. 

\section{Detector Simulation}
\label{Detector Simulation}
We performed full simulation including $e^+e^- \to \gamma Z$ and possible background processes.
The whole set of software programs used in this analysis is packaged as iLCSoft version v02-02\,\cite{iLCSoft}\,\cite{s3}\,\cite{s6}. 
Events were generated using Whizard 2.85\,\cite{s6} based on full tree-level helicity amplitudes for a given final state including non-resonant diagrams. Interactions of generated particles with the detector material are simulated with a full detector simulator based on GEANT4\,\cite{Geant4} using DD4hep (Detector Description for HEP)\,\cite{dd4hep}, which is the common detector geometry description for iLCSoft, including the 14 mrad crossing angle, IP smearing and offset depending on initial particles.
The event reconstruction programs are implemented as event processors in the framework of Marlin\,\cite{Marlin}. 
The event simulation for this analysis has been done at the center-of-mass energy of 250\,GeV. The assumed integrated luminosity is $\int Ldt = 900\,\mathrm{fb}^{-1}$ each for the two beam polarizations $(P_{e^-}, P_{e^+}) = (-0.8, +0.3)$ and $(+0.8, -0.3)$.
In our analysis, all particles are forced to be clustered into 2 jets and the jet with higher reconstructed energy is defined as ``jet 1” and the other as ``jet 2”.

\section{Signal Definition and Background}
\begin{figure}[ht]
  \centering
   \begin{tabular}[c]{cc}
    \includegraphics[width=0.8\columnwidth] {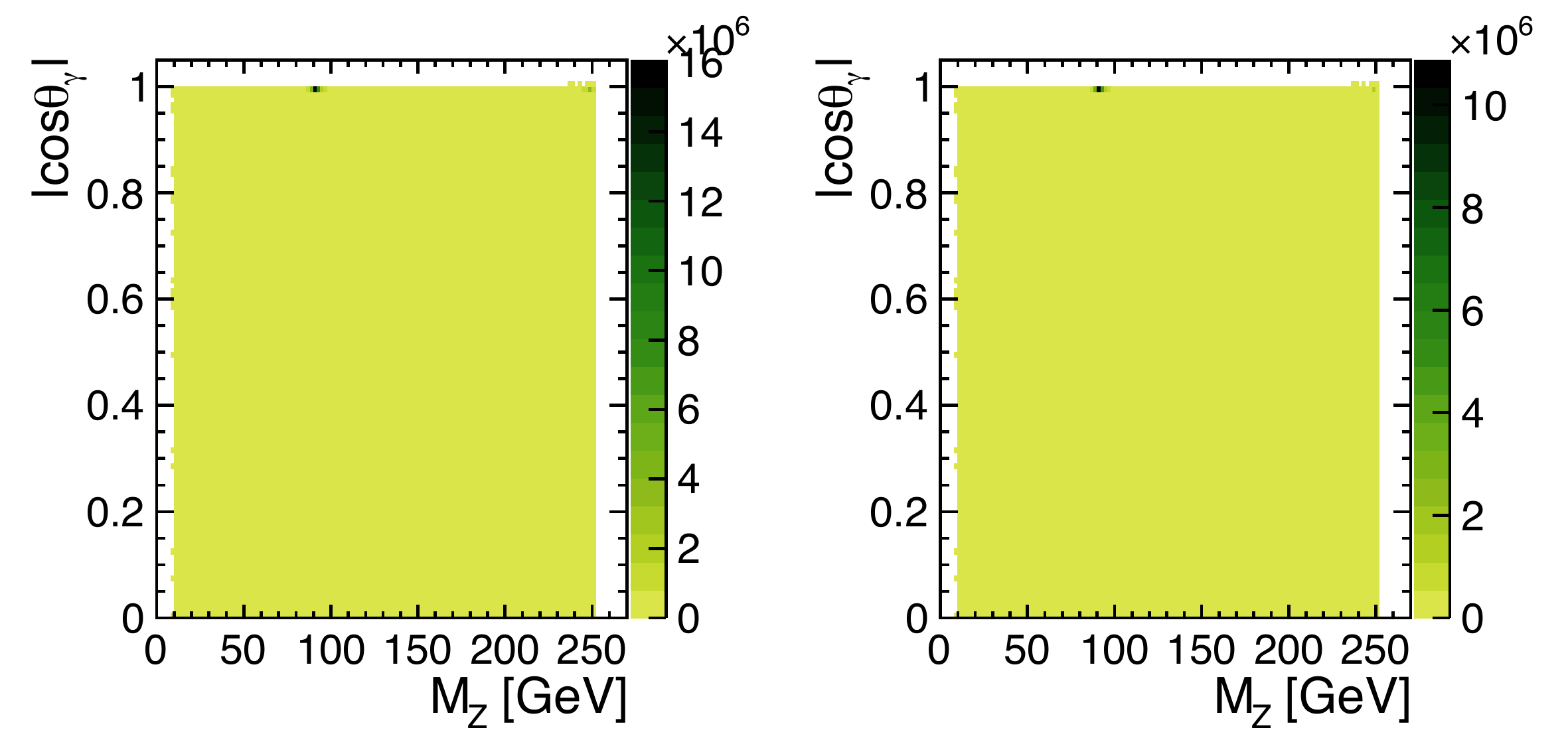}    
 \end{tabular}
    \caption{Photon angle and invariant mass of $Z$ boson distributions in the $e^+ e^- \to q\bar{q}$ samples. Each vertical axis is absolute value of cosine of polar angle of the signal photon. Left plot corresponds to $(P_{e^-}, P_{e^+}) = (-0.8, +0.3)$ case and right plot corresponds to $(+0.8, -0.3)$ case. Signal corresponds to $80\,\mathrm{GeV} < M_{q\bar{q}} $(MC truth)$ < 120\,\mathrm{GeV}$ region and most photons are going very forward.}
  \label{FA03}
\end{figure}

\begin{figure}[ht]
  \centering
   \begin{tabular}[c]{cc}
    \includegraphics[width=0.8\columnwidth] {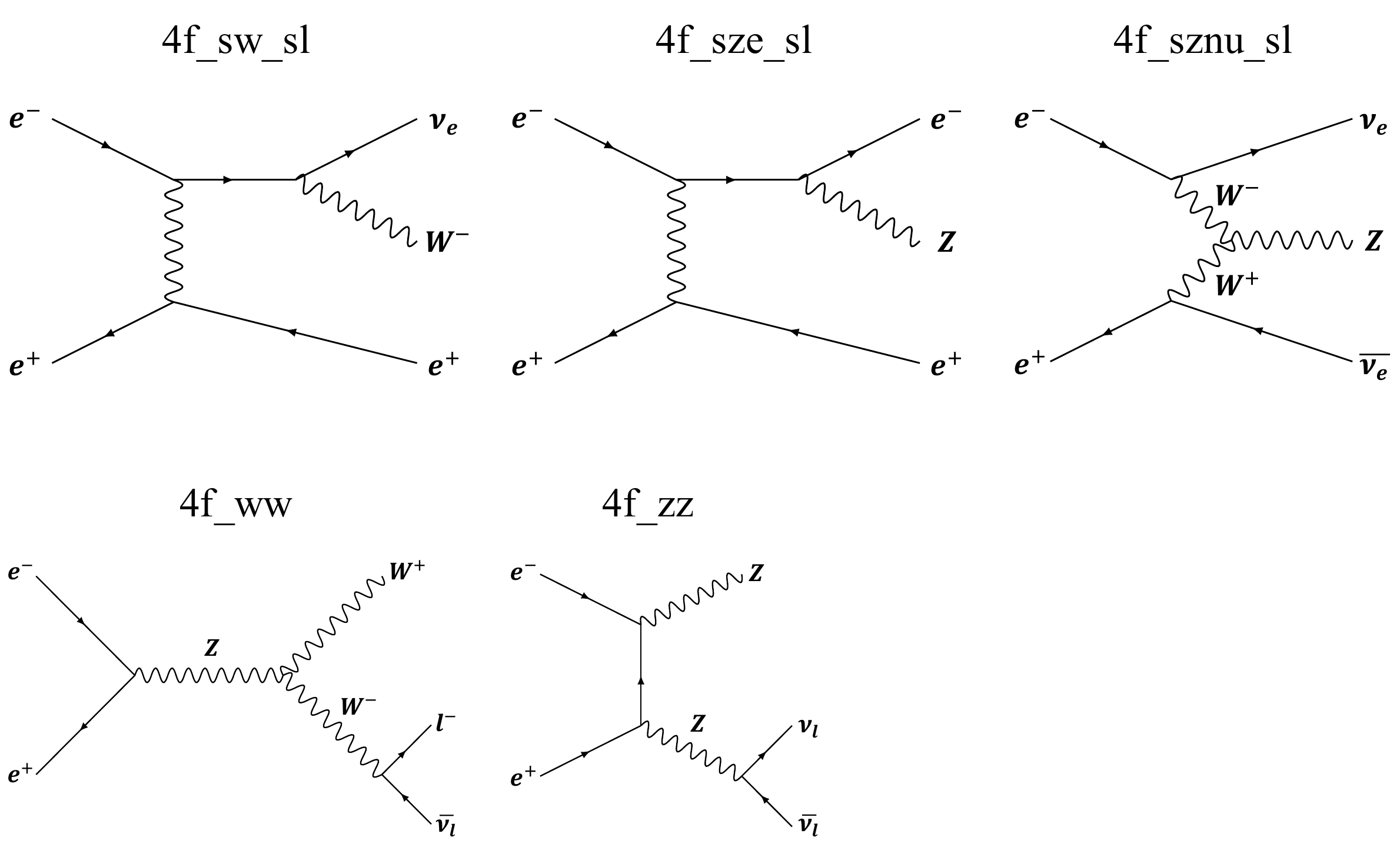}    
 \end{tabular}
    \caption{Potential background processes for the $e^+ e^- \to \gamma Z$ and $Z \to q\bar{q}$ process with abbreviated processes names.}
  \label{FA04}
\end{figure}

\begin{figure}[ht]
  \centering
   \begin{tabular}[c]{cc}
    \includegraphics[width=0.8\columnwidth] {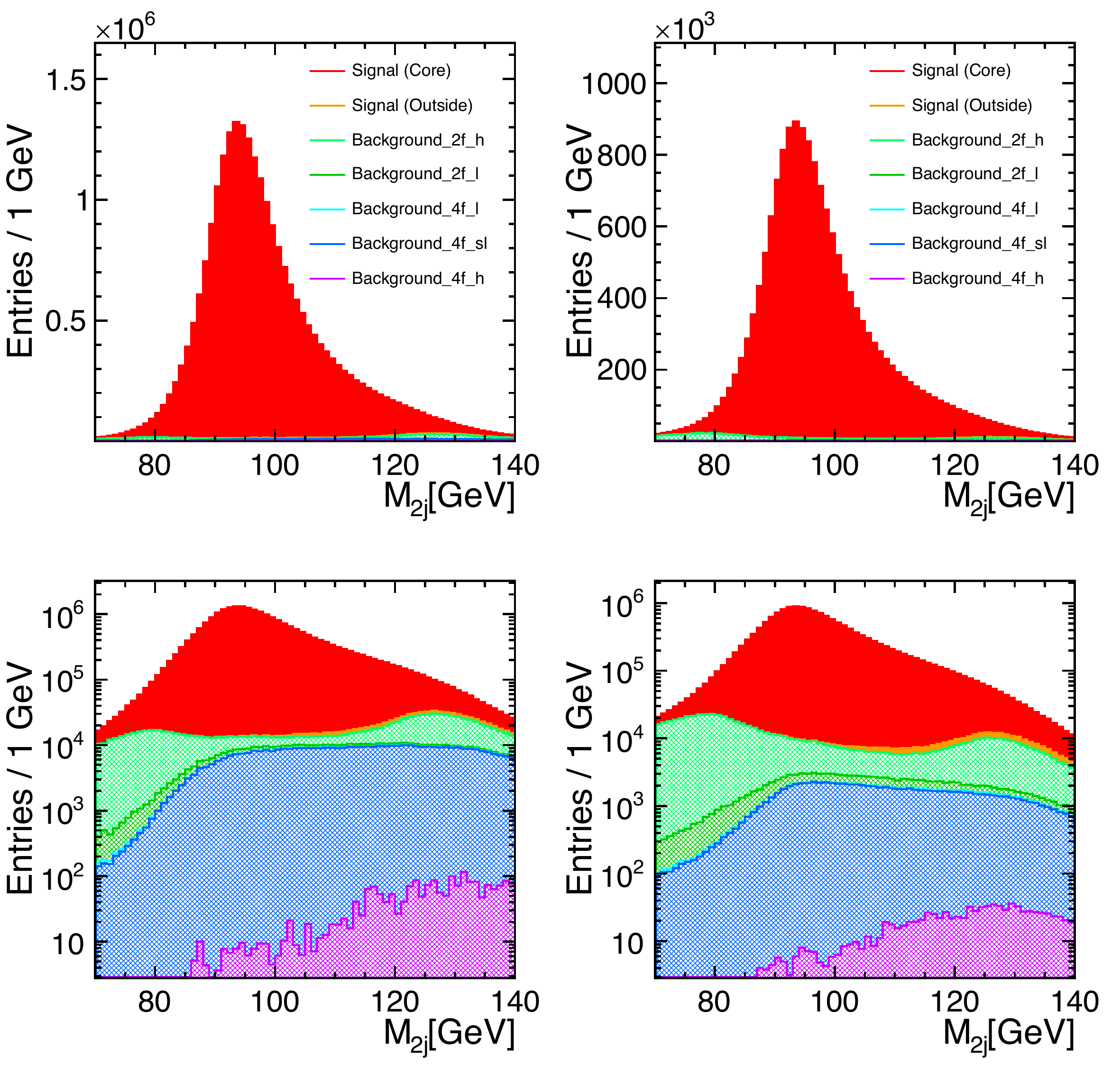}    
 \end{tabular}
    \caption{Stack plot of the invariant mass of 2-jet system $M_{2j}$ for the signal and background events for $(P_{e^-}, P_{e^+}) = (-0.8, +0.3)$ and $(+0.8, -0.3)$ polarization. Left plots are $(P_{e^-}, P_{e^+}) = (-0.8, +0.3)$ and right plots are $(+0.8, -0.3)$ polarization respectively after Cut 1 to 7. They are show in linear scale in top plots and in log scale in bottom plots.}
  \label{FA05}
\end{figure}

The signal for our analysis is $e^+ e^- \to \gamma Z$ and $Z \to q\bar{q}$ process satisfying $80\,\mathrm{GeV} < M_{q\bar{q}} $(MC truth)$ < 120\,\mathrm{GeV}$. As the radiative return photons are so collinear with the $e^-/e^+$ beam that they go into the beam pipe in most events (Fig.\,\ref{FA03}) and all the events which contain only 2 jets in the final state can be background, \textit{e.g.}\,those shown in Fig.\,\ref{FA04}. Our $e^+ e^- \to q\bar{q}$ samples contain events in which the $Z$ is far from the mass shell and we need to distinguish those events from the signal.\\
The considered background samples have a final state of two leptons ``2f\_l", two quarks ``2f\_h", four leptons ``4f\_l", two quarks and two leptons ``4f\_sl", and four quarks ``4f\_h". In order to suppress background events, background exclusion cuts are defined as follows by considering distributions of several useful observables.\\
\\
Cut 1\, $N_{\gamma (E> 50\,\mathrm{GeV})} = 0$\\ 
Cut 2\, $120\,\mathrm{GeV} < E_{vis}< 160\,\mathrm{GeV}$\\
Cut 3\, $|\cos \theta_{2j}| > 0.95$\\
Cut 4\, $N^{charged}_{J1} + N^{charged}_{J2} > 4$\\
Cut 5\, $N^{total}_{J1} + N^{total}_{J2} > 10$\\
Cut 6\, $50\,\mathrm{GeV} < M_{2j} < 160\,\mathrm{GeV}$\\
Cut 7\, $\cos \theta_{12} > -0.99$ or $\frac{E_{J1} - E_{J2}}{E_{J1} + E_{J2}} > 0.5$\\
\\
Here, $N_{\gamma (E> 50\,\mathrm{GeV})}$, $E_{vis}$, $|\cos \theta_{2j}|$, $N^{charged}_{J1}$, $N^{charged}_{J2}$,  $N^{total}_{J1}$, $N^{total}_{J2}$, $M_{2j}$,$\cos \theta_{12}$, $E_{J1}$, and $E_{J2}$ stands for the number of isolated photon with the energy more than 50 GeV, total visible energy, total momentum direction of the 2-jet system, the number of detected charged particles in jet 1, the number of detected charged particles in jet 2, the sum of the numbers of detected charged and neutral particles in jet 1, the sum of the numbers of detected charged and neutral particles in jet 2, the invariant mass of the 2-jet system, opening angle of two jets, energy of jet 1, and energy of jet 2, respectively.
Tables\,\ref{TA01} and \ref{TA02} show the luminosity normalized expected number of remaining events after each cut for $(P_{e^-}, P_{e^+}) = (-0.8, +0.3)$ and $(+0.8, -0.3)$ polarization, respectively at the ILC 250. A stack plot for the signal and background events after Cuts 1 through 7 is shown as a function of the reconstructed invariant mass of the 2-jet system $M_{2j}$ in Fig.\,\ref{FA05}.
\begin{table}[hbtp]
  \caption[Reduction table for signal and each background processes for $(P_{e^-}, P_{e^+}) = (-0.8, +0.3)$ polarization]{Reduction table for signal and each background processes for $(P_{e^-}, P_{e^+}) = (-0.8, +0.3)$ polarization, assuming $\int Ldt = 900\,\mathrm{fb}^{-1}$.}
  \label{TA01}
  \centering
  \begin{tabular}{lllllllll}
    \hline
    $\times 10^6$ events  & Signal & Signal (Core) & 2f\_l & 4f\_l & 4f\_sl & 4f\_h & 2f\_h & Bkg. Total\\
    \hline \hline
    Expected  &46.0& 32.5 & 12.7       &     9.34    & 17.2    & 15.1        & 23.6 & 78.1  \\
    Cut 1  &32.7& 31.1       & 10.1        & 5.96       & 16.0     & 14.8        & 21.6 & 68.3 \\
    Cut 2  &24.6& 24.4         & 2.55      & 1.46       & 3.22     & 0.00422  & 1.09& 8.32 \\
    Cut 3  &24.5& 24.4       & 1.93        & 0.366     & 0.526   & 0.00352  & 1.04 & 3.87 \\  
    Cut 4  &24.4& 24.3        &  0.299  & 0.0574 & 0.523   & 0.00352  & 1.00 & 1.88 \\   
    Cut 5  &24.3& 24.2      & 0.0651   & 0.0102 & 0.520   & 0.00352  & 0.977 & 1.58 \\   
    Cut 6  &24.2& 24.2        & 0.0571   & 0.00807  & 0.470  & 0.00210 & 0.694 & 1.23\\    
    Cut 7  &24.2& 24.1        & 0.0534   & 0.00647  & 0.463  & 0.00204  & 0.682 & 1.21\\   
    \hline
  \end{tabular}
\end{table}
\begin{table}[hbtp]
  \caption[Reduction table for signal and each background processes for $(P_{e^-}, P_{e^+}) = (+0.8, -0.3)$ polarization]{Reduction table for signal and each background processes for $(P_{e^-}, P_{e^+}) = (+0.8, -0.3)$ polarization, assuming $\int Ldt = 900\,\mathrm{fb}^{-1}$.}
  \label{TA02}
  \centering
  \begin{tabular}{lllllllll}
    \hline
    $\times 10^6$ events  & Signal & Signal (Core) & 2f\_l & 4f\_l & 4f\_sl & 4f\_h& 2f\_h & Bkg. Total\\
    \hline \hline
    Expected  &30.5& 21.6 & 9.84      & 5.50        & 2.56 & 1.41                    &10.6& 29.9  \\
    Cut 1  &21.7& 20.6       & 7.77       & 2.33       & 1.86 & 1.38                    &9.37& 22.7 \\
    Cut 2  &16.3& 16.2       & 1.83       & 0.378     & 0.370       & 0.00137      &1.04& 3.62  \\
    Cut 3  &16.3& 16.2       & 1.37       & 0.259     & 0.106       & 0.00124       &1.03& 2.77 \\  
    Cut 4  &16.2& 16.1       &  0.212    & 0.0357  & 0.104         & 0.00124     &0.985& 1.34 \\   
    Cut 5  &16.2& 16.1       & 0.0454   & 0.00603  & 0.102      & 0.00124      &0.958& 1.11 \\   
    Cut 6  &16.1& 16.0      & 0.0396   & 0.00468  & 0.0934     & 0.000986    &0.616& 0.754\\  
    Cut 7  &16.1& 16.0       & 0.0372  & 0.00320  & 0.0900    & 0.000967     &0.609& 0.740\\   
    \hline
  \end{tabular}
\end{table}

According to Tables\,\ref{TA01} and \ref{TA02}, signal selection efficiencies are $0.52678 \pm 0.00017$ and $0.52715 \pm 0.00016$ for $(P_{e^-}, P_{e^+}) = (-0.8, +0.3)$ and $(+0.8, -0.3)$ polarizations, respectively,
where we assumed the error on the efficiency is binomial. Background-to-signal ratios are 0.0499 and 0.0461 for $(P_{e^-}, P_{e^+}) = (-0.8, +0.3)$ and $(+0.8, -0.3)$ polarizations respectively after applying the seven cuts.

\section{Evaluation of the Error}
\label{Evaluation of the Error}
The error on $A_{LR}$ can be evaluated as below.
\begin{equation}
\begin{split}
\label{MA22}
\left(\frac{\Delta A_{LR}}{A_{LR}}\right)^2=\left(\frac{\Delta A_{LRobs}}{A_{LRobs}}\right)^2+\left(\frac{\Delta f}{f}\right)^2\\
\end{split}
\end{equation}
\begin{equation}
\begin{split}
\label{MA23}
\left(\frac{\Delta f}{f}\right)^2=\left(\frac{|P_{-}|(1+|P_{+}|)(1-|P_{+}|)}{(|P_{-}|+|P_{+}|)(1+|P_{-}||P_{+}|)}\right)^2 \left(\frac{\Delta |P_{-}|}{|P_{-}|}\right)^2+\left(\frac{|P_{+}|(1+|P_{-}|)(1-|P_{-}|)}{(|P_{-}|+|P_{+}|)(1+|P_{-}||P_{+}|)}\right)^2\left(\frac{\Delta |P_{+}|}{|P_{+}|}\right)^2.
\end{split}
\end{equation}
\begin{equation}
\begin{split}
\label{MA26}
\left(\frac {\Delta A_{LRobs}}{A_{LRobs}}\right)^2 &=\left(\frac{1}{2A_{LRobs}}\left(1-A_{LRobs}^2\right)\right)^2 \left(\left(\frac{\Delta \alpha}{\alpha}\right)^2 + \left(\frac{\Delta \beta}{\beta}\right)^2 + \left(\frac{\Delta N_{-}}{N_{-}}\right)^2 + \left(\frac{\Delta N_{+}}{N_{+}}\right)^2 \right).
\end{split}
\end{equation}

Here we assume the real experimental case of electron polarization $P_{-}$ being $|P_{-}| = 0.8$ and positron polarization $P_{+}$ being $|P_{+}| = 0.3$.
$A_{LRobs}$ is the asymmetry of the $Z$ production in $(P_{e^-}, P_{e^+}) = (-0.8, +0.3)$ and $(+0.8, -0.3)$ and $f$ is the polarization factor
\begin{equation}
\label{MA16}
f =  \frac{1+|P_{-}|| P_{+}|}{|P_{-}| + |P_{+}| },\\
\end{equation}
$N_{\pm}$ is the number of $Z$ production events for the eLpR and eRpL polarization, respectively,
and $\alpha$ and $\beta$ are the product of selection efficiency and luminosity for the eLpR and eRpL polarizations, respectively.

First, we estimated the statistical error
\begin{equation}
\label{MA14}
A_{LR} = 0.22810 \pm 0.00018\,(stat).
\end{equation}
This error is almost identical in the cases with and without background events, which confirms the number of background is sufficiently small.\\
The derived $A_{LR}$ value (\ref{MA14}) does not agree with the simulation setting which is 0.219.
This discrepancy was caused by the $e^+ e^- \to \gamma \to q\bar{q}$ diagram contamination in our sample.
However, we can cancel the deviation by taking appropriate $M_{q\bar{q}}$ range as our signal region.

We have so far been assuming that the polarization $|P_{-}|$ and $|P_{+}|$, selection efficiency $\eta$, and integrated luminosity $L$ have no errors. However, these have errors and the errors can cause further systematic error on $A_{LR}$. \\
When including the predicted polarization error of $\frac{\Delta |P_{-}|}{|P_{-}|} = \frac{\Delta |P_{+}|}{|P_{+}|} = 0.001$ into (\ref{MA23})\,\cite{whitepaper}, the total absolute error on $A_{LR}$ is estimated to be $0.000216$.\\
Next, errors on $\alpha$ and $\beta$ will be considered. Most of the error on $\alpha$ and $\beta$ are correlated because $\alpha$ and $\beta$ are evaluated in the same setup and we showed that the effect from this correlated part can be canceled out.
Therefore, the dominant source of the systematic error can be regarded as polarization error.
If $\frac{\Delta \alpha}{\alpha} = \frac{\Delta \beta}{\beta} = 0.00016$ (\textit{i.e.}\,0.016\%),
the total systematic error on $A_{LR}$ from polarization, selection efficiency, and luminosity is estimated to be $0.000174$, comparable to the statistical error $0.000178$.
In this case, total absolute error on $A_{LR}$ is $0.00025$, $8.8$ times better precision than that from the SLC ($0.00219$).

\section{Conclusion}
For the precision study at the ILC 250, measurement of $A_{LR}$ is important as it can constrain SMEFT parameters. We can use the $e^+ e^- \to \gamma Z$ process at the ILC to evaluate this observable. We performed a full simulation study including various background processes to assess by how much we can improve the precision of $A_{LR}$. 
We considered cut-based event selection to suppress various background processes.
The resultant statistical error on $A_{LR}$ is $1.8 \times 10^{-4}$.
When considering a relative polarization error as 0.001 for each polarization, the total absolute error on $A_{LR}$ is estimated to be $0.000216$.
If the integrated luminosity is adjusted to satisfy the product of luminosity and selection efficiency is the same for both polarization combinations, the correlated part of the error on this product would disappear.
Then we need to think about only uncorrelated parts. 
If $\frac{\Delta \alpha}{\alpha} = \frac{\Delta \beta}{\beta} = 0.00016$ (\textit{i.e.}\,0.016\%),
the total systematic error on $A_{LR}$ from polarization, selection efficiency, and luminosity is estimated to be $0.000174$, comparable to the statistical error $0.000178$.
In this case, the total absolute error on $A_{LR}$ is $0.00025$, $8.8$ times better precision than that from the SLC ($0.00219$).

\bigskip  \bigskip \begin{center} \begin{large}
                      \bf AKNOWLEDGEMENTS
\end{large} \end{center} 
We would like to thank the LCC generator working group and the ILD software working group for providing the simulation and reconstruction tools and producing the Monte Carlo samples used in this study.
This work has benefited from computing services provided by the ILC Virtual Organization, supported by the national resource providers of the EGI Federation and the Open Science GRID.
This work is supported in part by the Japan Society for the Promotion of Science under the Grants- in-Aid for Science Research 16H02173.



\end{document}